\documentclass[twocolumn]{aastex631}

\newcommand{\cluster}{SMACS\,J0723 }
\usepackage{amsmath}
\shorttitle{Discovery of a Low-mass Strong-lens System in SMACS\,J0723.3-7327}
\shortauthors{Deng et al. 2025}

\begin{document}

\title{Discovery of a Low-mass Strong-lens System in SMACS\,J0723.3-7327}

\author[0009-0009-9255-920X]{Limeng Deng}
\affiliation{Purple Mountain Observatory, Chinese Academy of Sciences,
              Nanjing, Jiangsu 210028, China}
\affiliation{School of Astronomy and Space Sciences, University of Science and Technology of China, 
              Hefei 230026, China}

\author[0000-0002-9063-698X]{Yiping Shu}
\affiliation{Purple Mountain Observatory, Chinese Academy of Sciences,
              Nanjing, Jiangsu 210028, China}
\correspondingauthor{Yiping Shu}
\email{yiping.shu@pmo.ac.cn}

\author[0000-0002-3779-9069]{Lei Wang}
\affiliation{Purple Mountain Observatory, Chinese Academy of Sciences,
              Nanjing, Jiangsu 210028, China}

\author[0000-0003-4211-851X]{Guoliang Li}
\affiliation{Purple Mountain Observatory, Chinese Academy of Sciences,
              Nanjing, Jiangsu 210028, China}

\author[0000-0001-6052-3274]{Gabriel Bartosch Caminha}
\affiliation{Technische Universität München, Physik-Department,
              James-Franck Str. 1, 85748 Garching, Germany}

\author[0009-0005-6876-7376]{Jiang Dong}
\affiliation{Purple Mountain Observatory, Chinese Academy of Sciences,
              Nanjing, Jiangsu 210028, China}
\affiliation{School of Astronomy and Space Sciences, University of Science and Technology of China, 
              Hefei 230026, China}

\author[0000-0001-8554-9163]{Zizhao He}
\affiliation{Purple Mountain Observatory, Chinese Academy of Sciences,
              Nanjing, Jiangsu 210028, China}

\author[0009-0000-8850-0250]{Han Qu}
\affiliation{Purple Mountain Observatory, Chinese Academy of Sciences,
              Nanjing, Jiangsu 210028, China}
\affiliation{School of Astronomy and Space Sciences, University of Science and Technology of China, 
              Hefei 230026, China}

\author[0000-0002-5155-5650]{Ruibiao Luo}
\affiliation{Purple Mountain Observatory, Chinese Academy of Sciences,
              Nanjing, Jiangsu 210028, China}

\begin{abstract}

We report the discovery of an intriguing, low-mass galaxy-scale strong-lens system in the SMACS\,J0723.3-7327 galaxy cluster. By modeling James Webb Space Telescope imaging and Very Large Telescope Multi-Unit Spectroscopic Explorer spectroscopic data, we find that the lens is cluster member galaxy at $z=0.397$ with an Einstein radius of 0\farcs424 $\pm$ 0\farcs012, stellar mass of $M_* = (3.3 \pm 0.8) \times 10^{10} M_\odot$, half-light radius of $\sim 1$ kpc, and central stellar velocity dispersion of $140 \pm 6$ km s$^{-1}$. This lens galaxy is one of the few strong lens galaxies known to date that have stellar mass as low as $M_* \sim 10^{10.5} M_\odot$, offering an exceptional opportunity to peek into the population of low-mass galaxies that has largely remained unexplored in the context of strong-lensing studies. This strong lens system can also assist in assessing the systematic uncertainty in the lens modeling of cluster member galaxies.

\end{abstract}
\keywords{
\href{https://astrothesaurus.org}{Strong gravitational lensing (1643)} ---
\href{https://astrothesaurus.org}{Galaxy formation (595)} ---
\href{https://astrothesaurus.org}{Galaxy evolution (594)}
}

\section{Introduction} \label{sec:intro}

Strong gravitational lensing is the phenomenon of forming multiple images of a background source object by the gravity of a foreground lens object. In particular, galaxy-scale strong-lens systems, where the lens objects are galaxies, are powerful tools for understanding galaxy formation and evolution. For example, it has been suggested that the lens galaxies studied generally prefer a Salpeter initial mass function \citep[IMF; e.g.,][]{Spiniello2011, Oguri2014} and their dark-matter fractions increase with stellar mass and velocity dispersion \citep[e.g.,][]{Auger2010b, Shu2015, Shajib2021}. Correlations between the central mass density profile of lens galaxies and galaxy properties such as stellar mass density, stellar mass, and redshift have been characterized \citep[e.g.,][]{Auger2010b, Bolton2012, 2013ApJ...777...97S, Shu2015, Li2018}, providing constraints on the impact of baryonic physics and mergers \citep[e.g.,][]{Nipoti2009, Velliscig2014, Sonnenfeld2014}. By measuring the shape and alignment of the dark-matter distribution and stellar mass distribution, studies have shown that dark-matter halos of those lens galaxies are typically rounder than the stellar mass distributions and the misalignment angles between dark matter and stars are generally small \citep[e.g.,][]{Koopmans2006, Bruderer2016, Shu2016}.

Nevertheless, such knowledge has been largely limited to massive galaxies because lens galaxies in currently known strong-lens samples typically have stellar mass of $10^{11}$--$10^{12} M_\odot$. On the other hand, stellar dynamical analyses of nearby galaxies ($z \lesssim 0.1$) have already reached down to the scale of dwarfs \citep[e.g.,][]{Salucci2019}. For instance, \citet{Cappellari2013} suggested that the dark-matter fraction reaches a minimum at a characteristic stellar mass of $M_* \sim 3 \times 10^{10} M_\odot$ and increases toward both low- and high-mass ends. The IMF is also found to vary with velocity dispersion in the sense that galaxies with lower velocity dispersions tend to have lighter IMF \citep[e.g. Chabrier or Kroupa,][]{Cappellari2013b, Li2017}. Low-mass lens galaxies ($M_* \lesssim 10^{10.5} M_\odot$) can hence provide highly complementary constraints, especially at high redshifts where spatially-resolved stellar kinematics observations remain challenging. 

\begin{figure*}[t!]
    \centering
    \includegraphics[width=0.96\textwidth]{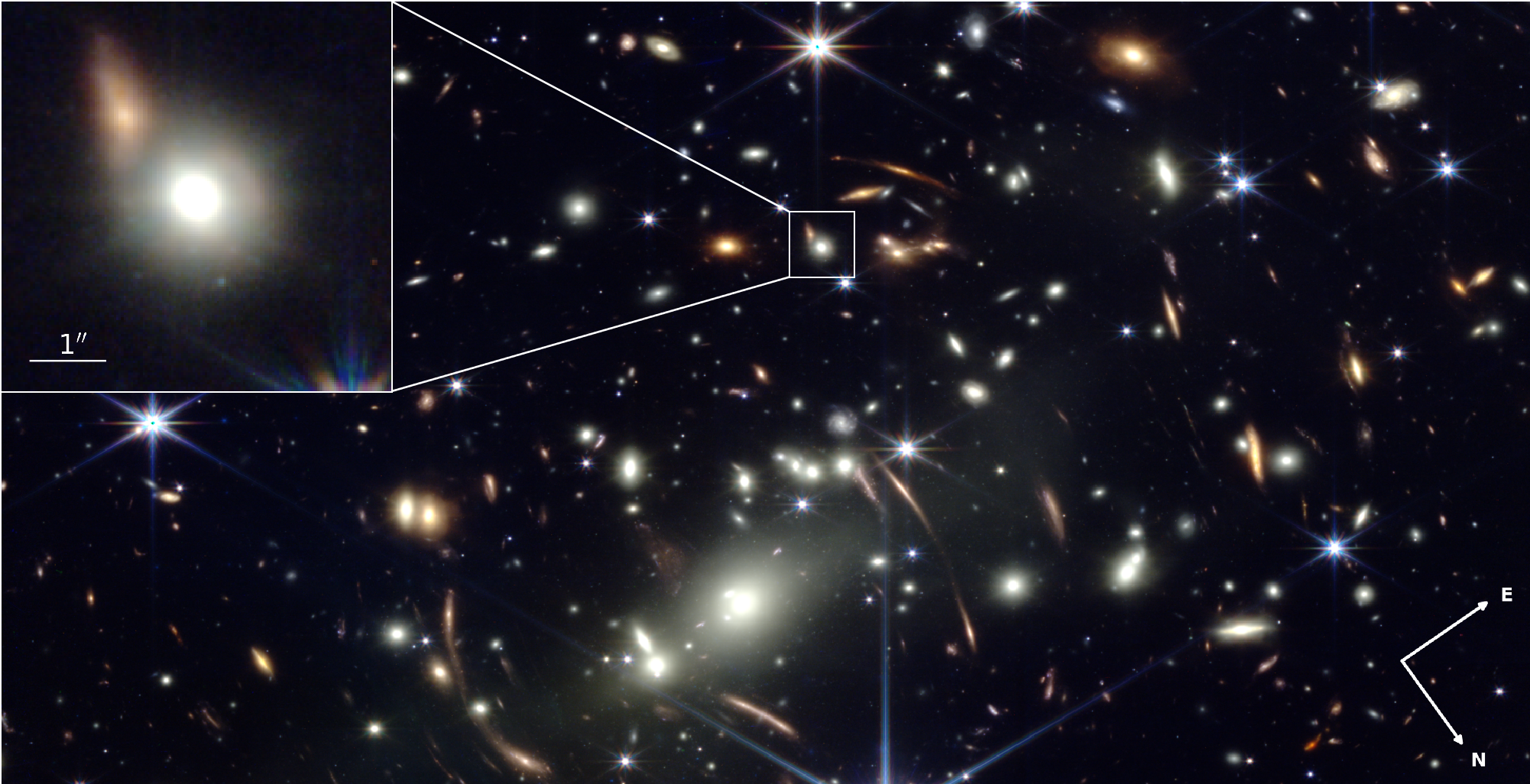}
    \caption{Color composite image of SMACS\,J0723 generated from JWST NIRCam data in six filters (F090W, F150W, F200W, F277W, F356W, and F444W). The inset figure provides an enlarged view of the strong-lens system studied in this work.}
    \label{fig:jwst}
\end{figure*}

A key obstacle in discovering low-mass lenses is the need for high angular resolution data. For typical lens and source redshifts of 0.5 and 1.5, the angular separations between multiple lensed images are on the order of $2^{\prime \prime}$ when the lensing mass is $10^{11.5} M_\odot$, which can be easily resolved in existing imaging data from wide-field surveys such as the Sloan Digital Sky Survey, Dark Energy Survey, Legacy Surveys, etc. When the lensing mass becomes $10^{10.5} M_\odot$, the angular separations decrease to $\approx$0\farcs7, significantly hindering the detectability. 

In this work, we will report a serendipitous discovery of a low-mass ($M_* \approx 10^{10.5} M_\odot$) strong-lens system from James Webb Space Telescope (JWST) observations of the SMACS\,J0723.3-7327 galaxy cluster (hereafter SMACS\,J0723). This Letter is organized as follows. We describe the imaging and spectroscopic data in Section~\ref{sec:Data} and present the strong-lens modeling procedures and results in Section~\ref{sec:Modeling}. Discussions and conclusions are provided in Section~\ref{sec:Discussion} and Section~\ref{sec:Conclusion}. All magnitudes are on the AB scale, and a standard concordance cosmology with $\Omega_m = 0.3$, $\Omega_{\Lambda} = 0.7$, and $h = 70$ km s$^{-1}$ Mpc$ ^{-1}$ is assumed.

\section{Data} \label{sec:Data}

\subsection{JWST Imaging Data}

The JWST Early Release Observations (ERO) of the SMACS\,J0723 galaxy cluster include broadband imaging by the NIRCam and MIRI as well as multi-object and slitless spectroscopy by the NIRSpec and NIRSS \citep[Programme ID: 2736,][]{2022ApJ...936L..14P}. In this work, we focused on the NIRCam imaging data in six filters covering the wavelength range from roughly 0.8 to 5.0 $\mu m$, i.e., F090W, F150W, F200W, F277W, F356W, and F444W. In each NIRCam filter, observations were split into nine subexposures (following the INTRAMODULEX dither pattern) with a total exposure time of $\approx$7537 s. Regarding data reduction, we made use of the UPdec-Webb data set constructed by \citet{2025ApJS..276...36W}, who reduced and coadded the NIRCam imaging data of SMACS\,J0723 using a customized algorithm---Up-sampling and Point-spread Function (PSF) Deconvolution Coaddition \citep{2022MNRAS.517..787W}. As demonstrated in \citet{2025ApJS..276...36W}, the UPdec-Webb images exhibit significant improvement in terms of photometry accuracy and faint source detectability compared to coadded images processed by the standard drizzle algorithm. The UPdec-Webb data set\footnote{Available at the National Astronomical Data Center: doi:10.12149/101436.} contains 11 imaging files per filter, which correspond to a directly coadded image (iter0) and 10 deconvolved images produced after different iterations of PSF deconvolution (iter1-iter10). In this work, we used the UPdec-Webb coadded images, which are supersampled by a factor of 2, resulting in a pixel scale of $0\farcs0155\,\mathrm{pixel}^{-1}$ in F090W, F150W, and F200W and $0\farcs031\,\mathrm{pixel}^{-1}$ in F277W, F356W, and F444W. 

A color composite image of the central region of SMACS\,J0723 constructed from the UPdec-Webb coadded images in all six NIRCam filters is presented in Figure~\ref{fig:jwst}. The galaxy-scale strong-lens system we discovered and analyzed in this work, denoted as SMACS\,J0723-SL, is located at R.A. = $110^\circ.8488458$, decl. = $-73^\circ.4606771$, about 31\farcs3 ($\approx$167 kpc) southeast from the brightest cluster galaxy (BCG). In this system, we clearly see an orange-ish question-mark structure surrounding the central galaxy, indicative of a strong-lensing effect. The radius of the structure is about 0\farcs8.

\subsection{MUSE Data Cube}

The \cluster cluster was also observed by the Multi-Unit Spectroscopic Explorer (MUSE) integral field spectrograph in 2019 under Programme ID: 0102.A-0718 (PI: A. Edge). In this work, we used the reduced data cube from \citet{2022A&A...666L...9C}, which has a 1 arcmin$^2$ field of view centered on the BCG and covers the wavelength range 4750\AA–-9350\AA. The spectral resolution is $\approx 2.4$\AA \ with a sampling of 1.25\AA \, per spectral pixel. According to the redshift catalog built from the MUSE data cube by \citet{2022A&A...666L...9C}, the lens galaxy in SMACS\,J0723-SL has a spectroscopic redshift of 0.3970, confirming it as a cluster member galaxy. In addition, the bright tail of the question-mark structure was also cataloged in \citet{2022A&A...666L...9C} with a spectroscopic redshift of 1.4792. The two different redshifts further support the strong-lensing interpretation. 

\section{Strong-lens Modeling} \label{sec:Modeling}

Encouraged by the imaging and spectroscopic evidence, we aim to obtain a sensible strong-lens model for SMACS\,J0723-SL to firmly establish its strong-lensing nature. Although strong-lens models of the \cluster cluster have been constructed by several teams \citep[e.g.,][]{2022A&A...666L...9C, Golubchik2022, Sharon2023, 2023ApJ...945...49M}, the lens galaxy in SMACS\,J0723-SL was either completely ignored or modeled simultaneously with all other member galaxies following a scaling relation, without using any constraint from the potential lensing features around it. In this work, we explicitly model the lens galaxy in SMACS\,J0723-SL. 

\begin{figure*}[ht!]
    \centering  
    \includegraphics[width=0.96\textwidth]{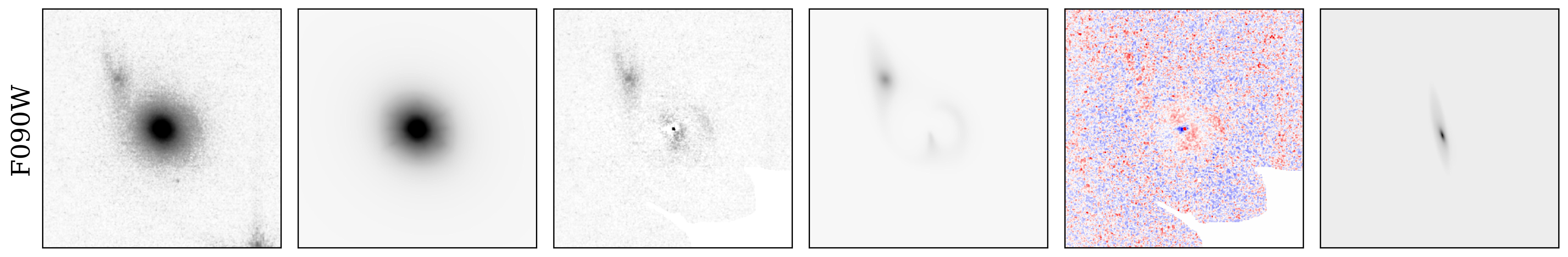} 
    \includegraphics[width=0.96\textwidth]{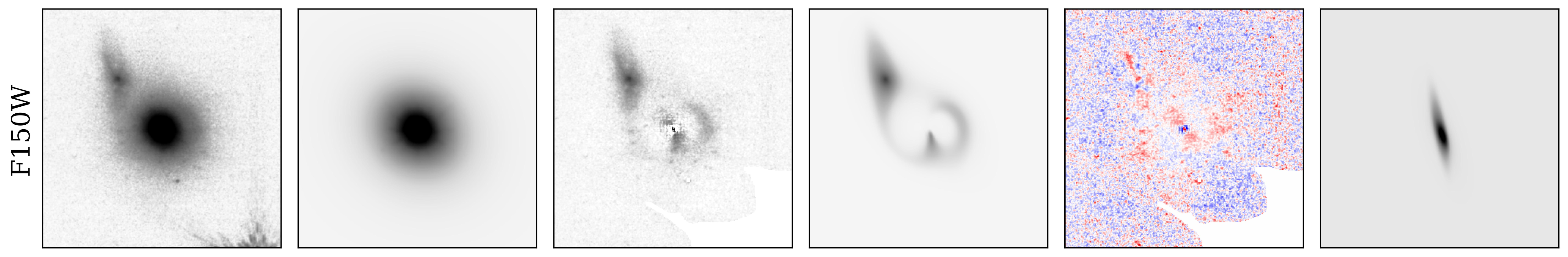} 
    \includegraphics[width=0.96\textwidth]{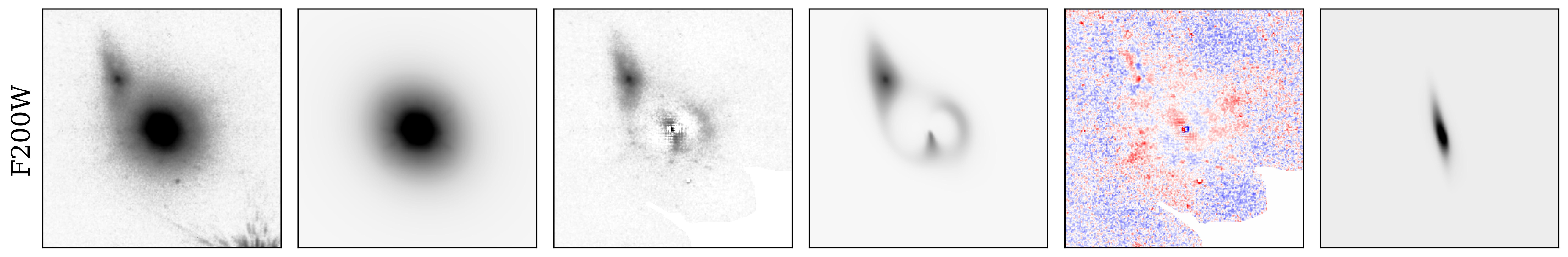} 
    \includegraphics[width=0.96\textwidth]{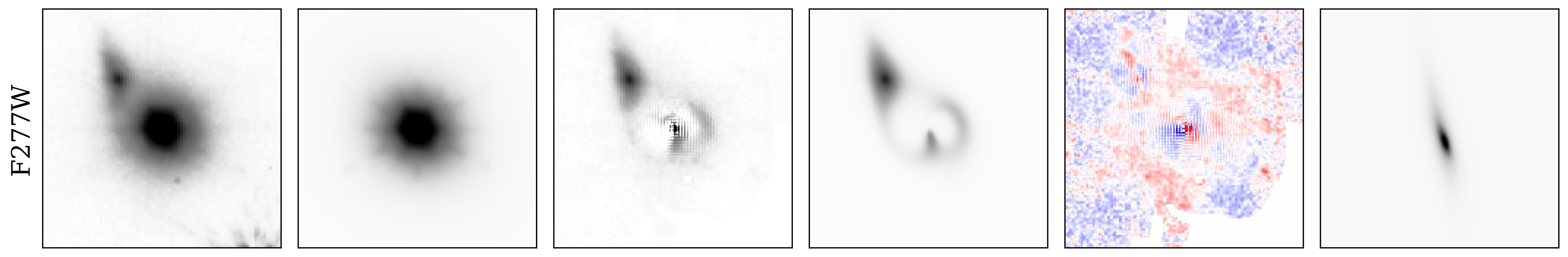} 
    \includegraphics[width=0.96\textwidth]{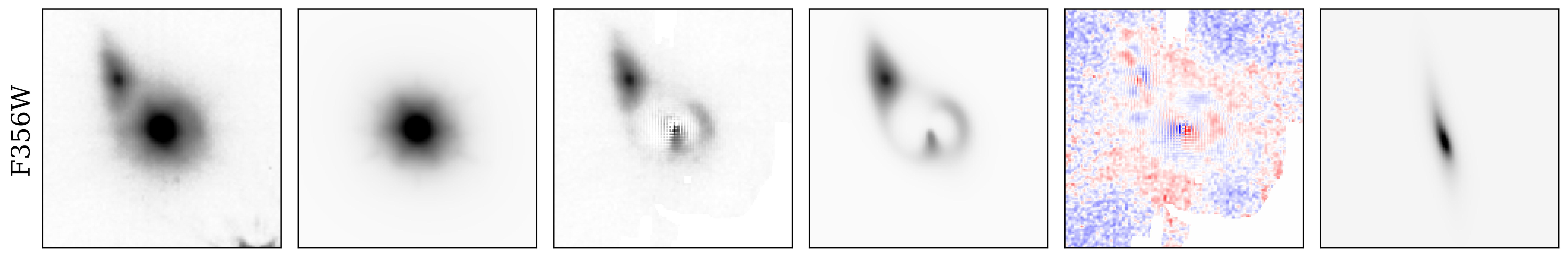} 
    \includegraphics[width=0.96\textwidth]{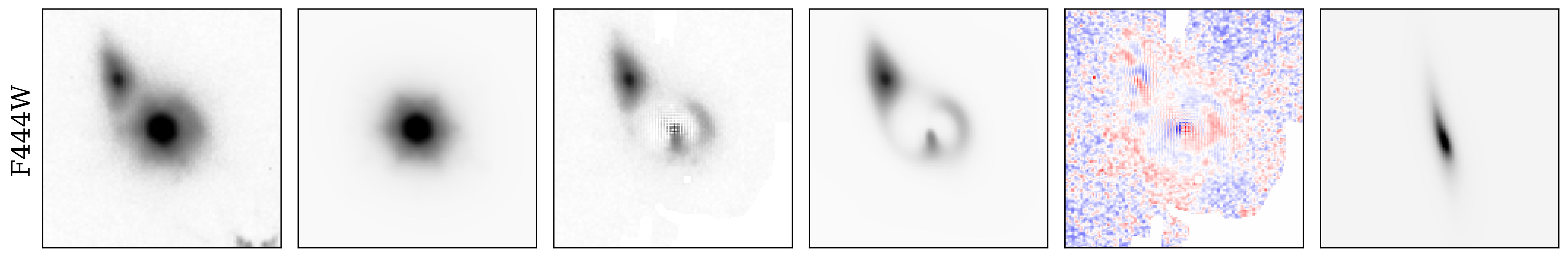} 
    \includegraphics[width=0.96\textwidth]{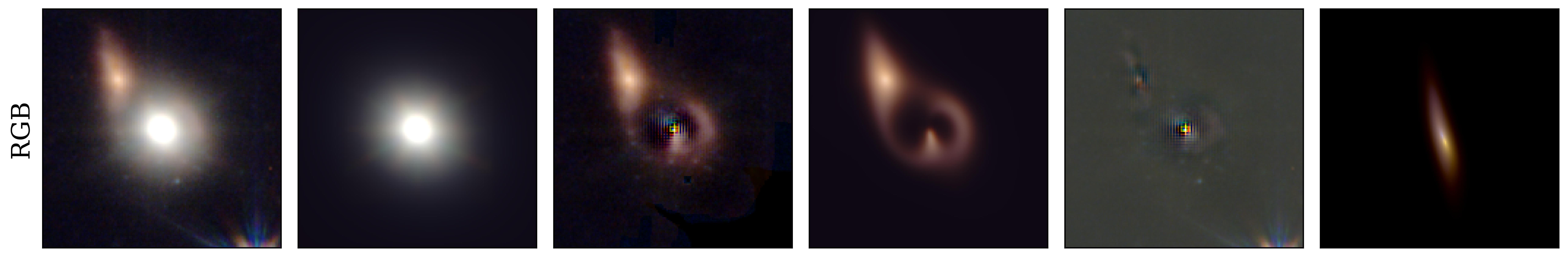} 
    \caption{Best-fitting results. Within each of the first six rows, from left to right, we show the data, lens light model, lens light-subtracted data, lensed image model, normalized residual, and source light model. The first six rows correspond to the six filters and the last row is a color composite equivalent generated by combining the six filters. }
    \label{fig:model}
\end{figure*} 

\begin{deluxetable*}{lccccccc}[!htbp]
    \tablecaption{Lens model Parameters\label{table:1} (Median and the 68\% Confidence Interval). }
    \tablehead{
    \colhead{$\theta_{\rm E}$ ($^{\prime\prime}$)} & 
    \colhead{$\gamma_{\rm EPL}$} & 
    \colhead{$\phi_{\rm EPL}$ (deg)} & 
    \colhead{$q_{\rm EPL}$} & 
    \colhead{$\gamma_{\rm shear}$} &
    \colhead{$\phi_{\rm shear}$ (deg)}
    }
    \startdata
    $0.424_{-0.012}^{+0.012}$ & $2.10_{-0.07}^{+0.03}$ & $-60.4_{-5}^{+1.8}$ & $0.87_{-0.04}^{+0.05}$ & $0.152_{-0.022}^{+0.018}$ & 
    $-20.7_{-0.5}^{+0.7}$ \\
    \enddata
\end{deluxetable*}

\begin{deluxetable*}{lccccccc}[!htbp]
\vspace{-1cm}
    \tablecaption{Lens light parameters\label{table:2}  (median and the 68\% confidence interval). }
    \tablehead{
    \colhead{Filter} & 
    \colhead{$m_{\rm AB}$} & 
    \colhead{$\Delta x$} & 
    \colhead{$\Delta y$} & 
    \colhead{$\phi$} & 
    \colhead{$q$} & 
    \colhead{$R_e$} & 
    \colhead{$n$} \\
    [-1.6ex]
    \colhead{} & 
    \colhead{} & 
    \colhead{(arcsec)} & 
    \colhead{(arcsec)} & 
    \colhead{(deg)} & 
    \colhead{} & 
    \colhead{(arcsec)} & 
    \colhead{} \\
    [-1.6ex]
    \colhead{(1)} & 
    \colhead{(2)} & 
    \colhead{(3)} & 
    \colhead{(4)} & 
    \colhead{(5)} & 
    \colhead{(6)} & 
    \colhead{(7)} & 
    \colhead{(8)}
    }
    \startdata
    F090W & $20.8526_{-0.0022}^{+0.0017}$ & $0.00405_{-0.00007}^{+0.00006}$ & $-0.01018_{-0.00021}^{+0.00010}$ & $-42.19_{-0.26}^{+0.18}$ & $0.8746_{-0.0020}^{+0.0014}$ & $0.2604_{-0.0010}^{+0.0011}$ & $4.948_{-0.015}^{+0.011}$ \\
    F150W & $19.9180_{-0.0014}^{+0.003}$ & $0.00405_{-0.00007}^{+0.00006}$ & $-0.01018_{-0.00021}^{+0.00010}$ & $-42.19_{-0.26}^{+0.18}$ & $0.8746_{-0.0020}^{+0.0014}$ & $0.2540_{-0.0012}^{+0.0008}$ & $4.616_{-0.04}^{+0.013}$ \\
    F200W & $19.6748_{-0.0007}^{+0.0008}$ & $0.00405_{-0.00007}^{+0.00006}$ & $-0.01018_{-0.00021}^{+0.00010}$ & $-42.19_{-0.26}^{+0.18}$ & $0.8746_{-0.0020}^{+0.0014}$ & $0.2498_{-0.0006}^{+0.0009}$ & $4.496_{-0.028}^{+0.03}$ \\
    F277W & $19.7811_{-0.003}^{+0.0024}$ & $0.00405_{-0.00007}^{+0.00006}$ & $-0.01018_{-0.00021}^{+0.00010}$ & $-42.19_{-0.26}^{+0.18}$ & $0.8746_{-0.0020}^{+0.0014}$ & $0.1998_{-0.0006}^{+0.0006}$ & $4.428_{-0.04}^{+0.029}$ \\
    F356W & $20.2936_{-0.0022}^{+0.003}$ & $0.00405_{-0.00007}^{+0.00006}$ & $-0.01018_{-0.00021}^{+0.00010}$ & $-42.19_{-0.26}^{+0.18}$ & $0.8746_{-0.0020}^{+0.0014}$ & $0.2072_{-0.0010}^{+0.0010}$ & $3.896_{-0.017}^{+0.023}$ \\
    F444W & $20.587_{-0.006}^{+0.006}$ & $0.00405_{-0.00007}^{+0.00006}$ & $-0.01018_{-0.00021}^{+0.00010}$ & $-42.19_{-0.26}^{+0.18}$ & $0.8746_{-0.0020}^{+0.0014}$ & $0.1935_{-0.0024}^{+0.0022}$ & $3.587_{-0.011}^{+0.016}$ \\
    \enddata
    \tablecomments{Column (2) is the integrated magnitude. Columns (3)--(4) are the $x$- and $y$-coordinates of the S\'ersic component relative to the cutout center. Columns (5)--(8) correspond to the position angle of the major axis (east from north), minor-to-major axis ratio, half-light radius, and S\'ersic index.}
\end{deluxetable*}

\begin{deluxetable*}{lcccccccc}[!htbp]
\vspace{-1cm}
    \tablecaption{Source light parameters\label{table:3} (median and the 68\% confidence interval). }
    \tablehead{
    \colhead{Filter} & 
    \colhead{$m_{\rm AB}$} & 
    \colhead{$\Delta x$} & 
    \colhead{$\Delta y$} & 
    \colhead{$\phi$} & 
    \colhead{$q$} & 
    \colhead{$R_e$} & 
    \colhead{$n$} & 
    \colhead{$\mu$} \\
    [-1.6ex]
    \colhead{} & 
    \colhead{} & 
    \colhead{(arcsec)} & 
    \colhead{(arcsec)} & 
    \colhead{(deg)} & 
    \colhead{} & 
    \colhead{(arcsec)} & 
    \colhead{} & 
    \colhead{} \\
    [-1.6ex]
    \colhead{(1)} & 
    \colhead{(2)} & 
    \colhead{(3)} & 
    \colhead{(4)} & 
    \colhead{(5)} & 
    \colhead{(6)} & 
    \colhead{(7)} & 
    \colhead{(8)} & 
    \colhead{(9)}
    }
    \startdata
    \multicolumn{9}{l}{\textbf{Component 1}} \\ 
    F090W & $26.919_{-0.029}^{+0.06}$ & $-0.087_{-0.010}^{+0.014}$ & $0.368_{-0.021}^{+0.016}$ & $-75.8_{-2.4}^{+2.6}$ & $0.331_{-0.007}^{+0.005}$ & $0.102_{-0.003}^{+0.003}$ & $1.495_{-0.015}^{+0.04}$ & 7.8 \\
    F150W & $25.36_{-0.04}^{+0.06}$ & $-0.087_{-0.010}^{+0.014}$ & $0.368_{-0.021}^{+0.016}$ & $-75.8_{-2.4}^{+2.6}$ & $0.331_{-0.007}^{+0.005}$ & $0.1230_{-0.0019}^{+0.0016}$ & $1.800_{-0.009}^{+0.007}$ & 8.1 \\
    F200W & $25.12_{-0.05}^{+0.09}$ & $-0.087_{-0.010}^{+0.014}$ & $0.368_{-0.021}^{+0.016}$ & $-75.8_{-2.4}^{+2.6}$ & $0.331_{-0.007}^{+0.005}$ & $0.1143_{-0.004}^{+0.0019}$ & $1.548_{-0.009}^{+0.04}$ & 8.0 \\
    F277W & $24.86_{-0.04}^{+0.10}$ & $-0.087_{-0.010}^{+0.014}$ & $0.368_{-0.021}^{+0.016}$ & $-75.8_{-2.4}^{+2.6}$ & $0.331_{-0.007}^{+0.005}$ & $0.0997_{-0.005}^{+0.0021}$ & $1.692_{-0.024}^{+0.062}$ & 7.7 \\
    F356W & $24.76_{-0.05}^{+0.08}$ & $-0.087_{-0.010}^{+0.014}$ & $0.368_{-0.021}^{+0.016}$ & $-75.8_{-2.4}^{+2.6}$ & $0.331_{-0.007}^{+0.005}$ & $0.093_{-0.004}^{+0.003}$ & $1.620_{-0.008}^{+0.007}$ & 7.7 \\
    F444W & $24.71_{-0.03}^{+0.06}$ & $-0.087_{-0.010}^{+0.014}$ & $0.368_{-0.021}^{+0.016}$ & $-75.8_{-2.4}^{+2.6}$ & $0.331_{-0.007}^{+0.005}$ & $0.0940_{-0.0020}^{+0.0022}$ & $1.656_{-0.027}^{+0.06}$ & 7.7 \\
    \hline
    \multicolumn{9}{l}{\textbf{Component 2}} \\ 
    F090W & $27.05_{-0.03}^{+0.05}$ & $-0.121_{-0.012}^{+0.014}$ & $0.453_{-0.017}^{+0.008}$ & $-82.4_{-2.0}^{+2.2}$ & $0.158_{-0.006}^{+0.006}$ & $0.133_{-0.007}^{+0.005}$ & $0.194_{-0.008}^{+0.006}$ & 8.6 \\
    F150W & $25.757_{-0.027}^{+0.04}$ & $-0.121_{-0.012}^{+0.014}$ & $0.453_{-0.017}^{+0.008}$ & $-82.4_{-2.0}^{+2.2}$ & $0.158_{-0.006}^{+0.006}$ & $0.1490_{-0.0022}^{+0.0024}$ & $0.270_{-0.016}^{+0.016}$ & 8.6 \\
    F200W & $25.634_{-0.021}^{+0.026}$ & $-0.121_{-0.012}^{+0.014}$ & $0.453_{-0.017}^{+0.008}$ & $-82.4_{-2.0}^{+2.2}$ & $0.158_{-0.006}^{+0.006}$ & $0.147_{-0.004}^{+0.005}$ & $0.409_{-0.018}^{+0.03}$ & 8.2 \\
    F277W & $25.229_{-0.022}^{+0.026}$ & $-0.121_{-0.012}^{+0.014}$ & $0.453_{-0.017}^{+0.008}$ & $-82.4_{-2.0}^{+2.2}$ & $0.158_{-0.006}^{+0.006}$ & $0.1425_{-0.003}^{+0.0024}$ & $0.591_{-0.016}^{+0.04}$ & 8.0 \\
    F356W & $25.045_{-0.022}^{+0.03}$ & $-0.121_{-0.012}^{+0.014}$ & $0.453_{-0.017}^{+0.008}$ & $-82.4_{-2.0}^{+2.2}$ & $0.158_{-0.006}^{+0.006}$ & $0.1381_{-0.0027}^{+0.0028}$ & $0.640_{-0.015}^{+0.04}$ & 7.8 \\
    F444W & $24.957_{-0.029}^{+0.05}$ & $-0.121_{-0.012}^{+0.014}$ & $0.453_{-0.017}^{+0.008}$ & $-82.4_{-2.0}^{+2.2}$ & $0.158_{-0.006}^{+0.006}$ & $0.1277_{-0.003}^{+0.0028}$ & $0.756_{-0.019}^{+0.016}$ & 7.6 \\
    \enddata
    \tablecomments{The first eight columns are arranged in the same order as Table~\ref{table:2}. Column (9) corresponds to the lensing magnification. }
\end{deluxetable*}

\vspace{-7em}
\subsection{Model Setup} \label{subsec:3.1}

For the lens galaxy in SMACS\,J0723-SL, we used an elliptical power-law (EPL) profile to model its projected total-mass distribution, which is parameterized as:
\begin{equation}
    \Sigma (x,y) = \Sigma_{\rm crit} \frac{3-\gamma_{\rm EPL} }{2} \left ( \frac{\theta_{\rm E} }{\sqrt{qx^{2}+y^{2}/q  } }  \right ) ^{\gamma_{\rm EPL} -1} ,
\end{equation}
where $\theta_{\rm E}$ is the Einstein radius, $\gamma_{\rm EPL}$ is the power-law slope, $q$ is the minor-to-major axis ratio, and $x$ and $y$ are defined in a coordinate system aligned with the major and minor axes of the lens mass distribution. $\Sigma_{\rm crit}$ is the critical density defined as $\Sigma_{\rm crit}=\frac{c^{2}}{4\pi G} \frac{D_{S}}{D_{L}D_{LS}}$, where $D_L$, $D_S$, and $D_{LS}$ are angular diameter distances of the lens, the source, and from the lens to the source, respectively.

To account for the contribution from the cluster to the lensing potential, we utilized the lens model of \cluster constructed using the JWST ERO data by \citet[][hereafter C22]{2022A&A...666L...9C}, or more specifically, their convergence and shear maps (100 each, corresponding to 100 Monte Carlo realizations). In principle, we should directly adopt C22 results as the external convergence and external shear in our model. However, only the shear strength map was available in C22 but not the shear position angle. We thus had to optimize for the external shear, the lensing potential of which is parameterized as 
\begin{equation}
    \psi_{\rm shear} = -\frac{1}{2} \gamma _{shear} (x^{2} + y^2) \cos 2\left (\phi - \phi _{shear} \right ) ,
\end{equation}
where $\gamma_{\rm shear}$ and $\phi_{\rm shear}$ are the strength and position angle of the external shear. 

Regarding the external convergence, we selected a $5^{\prime \prime}  \times 5^{\prime \prime}$ region centered on SMACS\,J0723-SL from the C22 convergence maps. We found that the mean convergence in this region is $\sim 0.4$ and the spatial variation is $\sim 0.017$. We therefore assumed the external convergence can be approximated as a constant $\kappa_{\rm ext}$, i.e., a mass sheet. We computed the mean convergence in the selected region for each realization and adopted the median and standard deviation of the 100 mean convergence, $0.40$ and $0.018$, as the median and $1\sigma$ uncertainty for $\kappa_{\rm ext}$. In our modeling process, we then fixed $\kappa_{\rm ext}$ to $0.40$. To account for the impact of the $\kappa_{\rm ext}$ uncertainty on the other lens parameters, we actually made use of the mass-sheet degeneracy \citep[MSD;][]{1985ApJ...289L...1F, 2013A&A...559A..37S, 2024MNRAS.533..795K}. According to the MSD, the following two sets of lensing potentials give identical lensed imaging signals, 
\begin{equation}
    \kappa^{\prime} (x, y) = \lambda \kappa (x, y )+(1-\lambda),
\end{equation}
when the source plane coordinate is also rescaled as $\vec{\beta} = \lambda \vec{\beta}$. Therefore, $\theta_{\rm E}$ and $\gamma_{\rm shear}$ will change with $\kappa_{\rm ext}$ as 
\begin{equation}
    \frac{\delta \theta _{\rm E}}{\theta _{\rm E}}= \frac{\delta \kappa _{\rm ext}}{(\gamma_{\rm EPL} - 1) \; (1-\kappa _{\rm ext})},
\end{equation}
\begin{equation}
     \frac{\delta \gamma_{\rm shear}}{\gamma_{\rm shear}}=\frac{\delta \kappa _{\rm ext}}{1-\kappa _{\rm ext}}.
\end{equation}

We used the S\'ersic profile \citep{1963BAAA....6...41S} to model the light distributions of the lens galaxy and source galaxy. The number of S\'ersic components was determined by the data. In each filter, we found that a single S\'ersic was sufficient for the lens galaxy, while two were needed for the source galaxy, which appears to have a relatively compact component and an extended component. The PSF models were derived from the (unsaturated) star closest to the lensing system using the method developed by \citet{2023arXiv230814065N}.

\begin{figure*}[t!]
    \centering
    \includegraphics[width=0.48\textwidth]{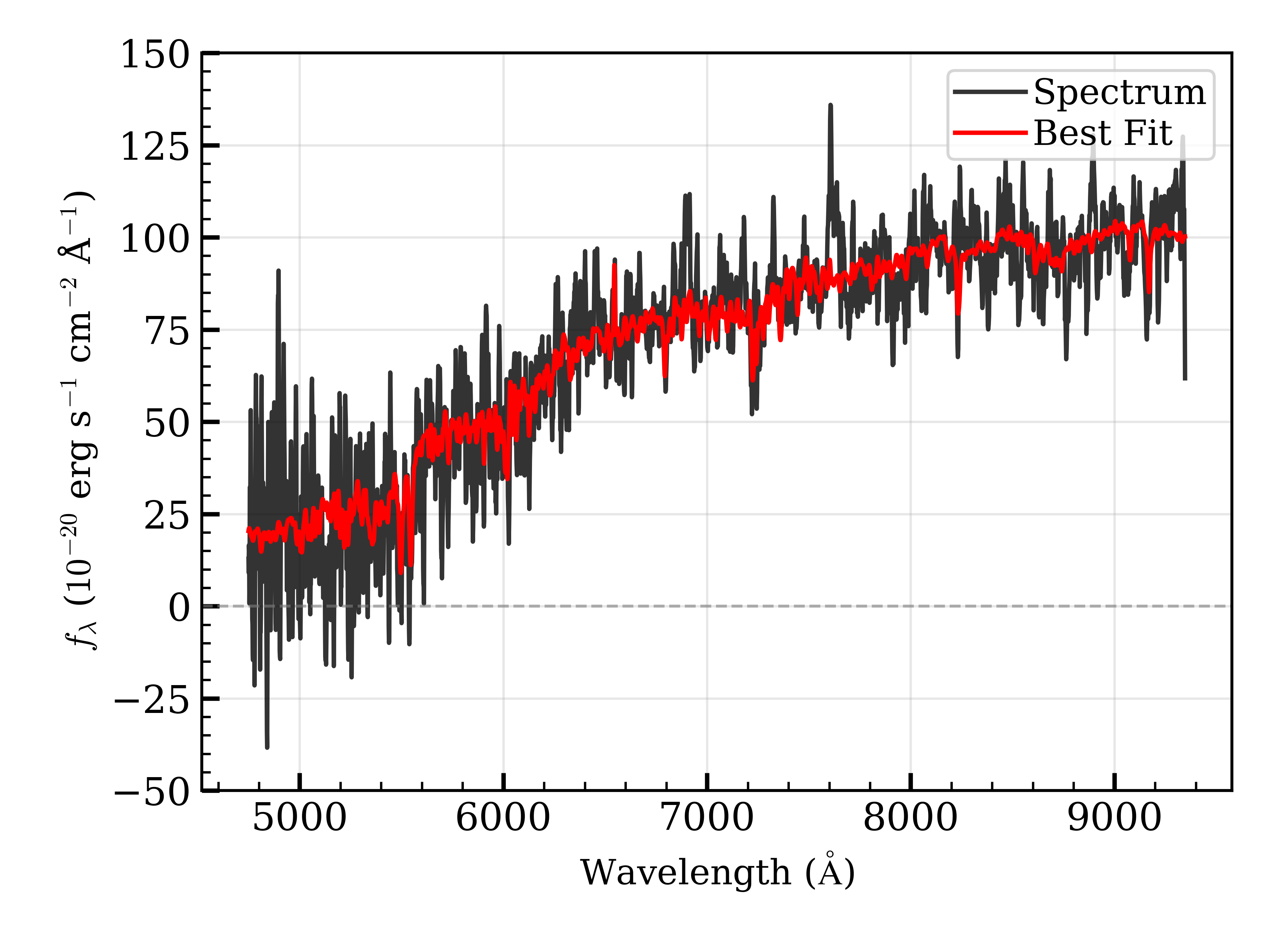}
    \hfill
    \includegraphics[width=0.48\textwidth]{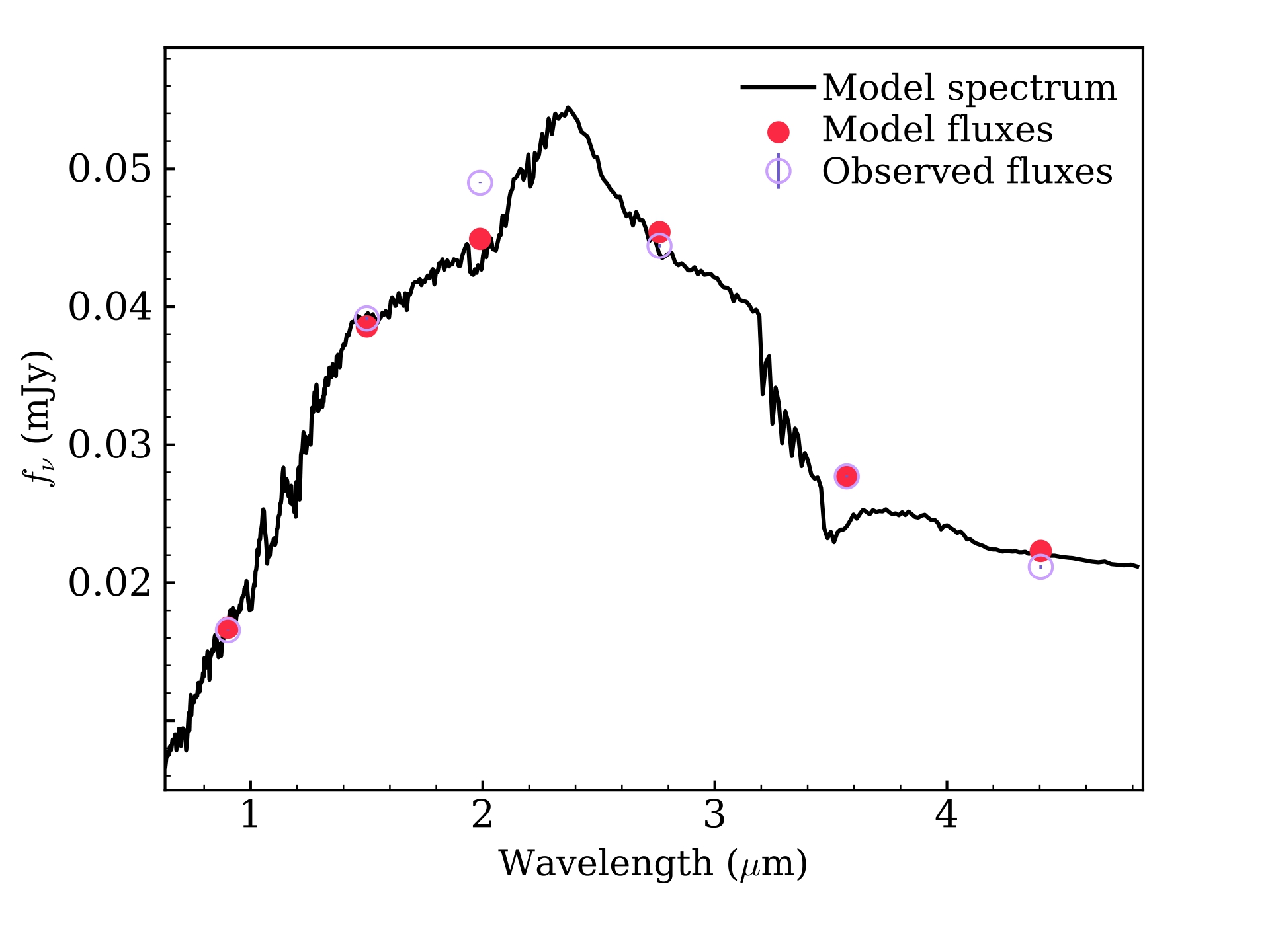}
    \caption{\emph{Left}: {\tt pPXF} velocity dispersion fitting result. The black line corresponds to the observed spectrum (smoothed for illustration purposes) and the red line corresponds to the best fit. \emph{Right}: {\tt CIGALE} SED fitting result. Open and filled circles correspond to the observed and model-predicted photometry.}
    \label{fig:velocity}
\end{figure*}

We simultaneously modeled the imaging data from all six filters. For the three shorter wavelength filters (i.e., F090W, F150W, and F200W), cutouts of $325 \times 325$ pixels centered on SMACS\,J0723-SL were used, while cutouts of $163 \times 163$ pixels were used for the three longer wavelength filters (i.e., F277W, F356W, and F444W). Each cutout is therefore roughly $5^{\prime \prime}$ across. 
In the modeling process, we assumed that the lens light distributions in the six filters have the same center, axis ratio, and position angle. For the source galaxy, each of the two S\'ersic models also has the same center, axis ratio, and position angle across the six filters. We further fixed the lens mass center to the lens light center. In total, our model contained 54 nonlinear parameters that need to be constrained, including 4 from the EPL model, 2 from the external shear model, 16 from the lens light models, and 32 from the source light models. The parameter inference was done using the open-source package {\tt lenstronomy} \citep{2018PDU....22..189B}. 

\subsection{Modeling Results\label{subsec:3.2}}

Our modeling results are presented in Figure~\ref{fig:model} and Tables~\ref{table:1}, \ref{table:2}, and \ref{table:3}. It can be seen that our model successfully reproduced the imaging data, especially the question-mark structures. The reduced $\chi^2$ value is 0.9976. The lens mass distribution is slightly steeper than isothermal (i.e., $\gamma=2$), and is well aligned with its light distribution in terms of both the axis ratio and position angle. We thus confirmed that SMACS\,J0723-SL is a genuine strong-lens system. Furthermore, we found that the lens galaxy has an exceptionally small Einstein radius of $0^{\prime \prime}.424 \pm 0^{\prime \prime}.012$. We note that this Einstein radius is what the lens galaxy would have if it were in the field (i.e., without considering the contribution from the cluster). The quoted uncertainty includes contributions from both the data noise and uncertainty on $\kappa_{\rm ext}$. As will be shown in Section \ref{sec:Discussion}, the small Einstein radius makes SMACS\,J0723-SL immediately stand out from other previously known strong lenses. 

The S\'ersic indices of the lens are about 4--5 in all six filters, suggesting that the lens is an early-type galaxy. The half-light radii of the lens are progressively smaller toward redder wavelength, starting from $\approx 0^{\prime \prime}.26$ (1.39 kpc) in F090W to $\approx 0^{\prime \prime}.19$ (1.02 kpc) in F444W. This trend is consistent with the inside-out growth scenario.

The two components of the source galaxy are separated by $\approx 0^{\prime \prime}.09$ (0.761 kpc) in the source plane and have distinct S\'ersic indices (both smaller than $n=2$). The brighter component, i.e., Component 1, is slightly more compact and less elongated compared to the fainter component. The lensing magnifications for the two components are $\sim 7.8$ and $\sim 8.1$, and they do not vary significantly across the six filters.


\section{Discussion} \label{sec:Discussion}

According to our best-fit model, the total enclosed mass of the lens galaxy in SMACS\,J0723-SL within the isodensity ellipse of semimajor axis $\theta_{\rm E}/\sqrt{q_{\rm EPL}}$ (2.43 kpc) and semiminor axis $\sqrt{q_{\rm EPL}} \theta_{\rm E}$ (2.11 kpc) is given by 
\begin{equation}
\begin{aligned}
    M_{\rm lensing} &= \pi \theta_{\rm E}^2 \Sigma_{\rm crit} \\
    &= (3.7 \pm 0.2) \times 10^{10} M_\odot.
\end{aligned}
\end{equation}
To estimate the stellar mass of the lens galaxy, we utilized {\tt CIGALE}, a public tool designed for fitting the spectral energy distribution (SED) of galaxies \citep{2019A&A...622A.103B}. We chose the single stellar population (SSP) library from \citet{2003MNRAS.344.1000B} and assumed a double exponential star formation history (the {\tt sfh2exp} module). The Salpeter stellar IMF \citep{1955ApJ...121..161S} was employed, and the metallicity is fixed to $Z = 0.02$. We used the {\tt dustatt\_modified\_starburst} module, a modification and extension of the \citet{2000ApJ...533..682C} attenuation law, for characterizing the dust attenuation. We also included dust emission using the {\tt casey2012} module, which is based on the dust emission model from \citet{2012MNRAS.425.3094C}. We used this model to fit for the observed lens galaxy SED in six NIRCam filters, and the fitting result is shown in Figure~\ref{fig:velocity}. We found a total stellar mass of $(3.3 \pm 0.8) \times 10^{10} M_\odot$. The projected dark-matter fraction within the half-light radius is thus $\approx (12 \pm 22) \%$. We further extracted the spectrum of a $0^{\prime \prime}.6 \times 0^{\prime \prime}.6$ region centered on the lens galaxy from the MUSE data cube and used the {\tt pPXF} package \citep{2004PASP..116..138C, 2017MNRAS.466..798C, 2023MNRAS.526.3273C} to derive the central stellar velocity dispersion of the lens galaxy. For this analysis, we utilized the E-MILES stellar population synthesis models \citep{2016MNRAS.463.3409V}, which provide UV-extended templates spanning the wavelength range of 1680--50000\AA. The fitting result is shown in Figure~\ref{fig:velocity}, and the lens galaxy was found to have a velocity dispersion of $\sigma_* = 140 \pm 6~\mathrm{km~s^{-1}}$. 

\begin{figure}[t!]
    \centering
    \includegraphics[width=0.48\textwidth]{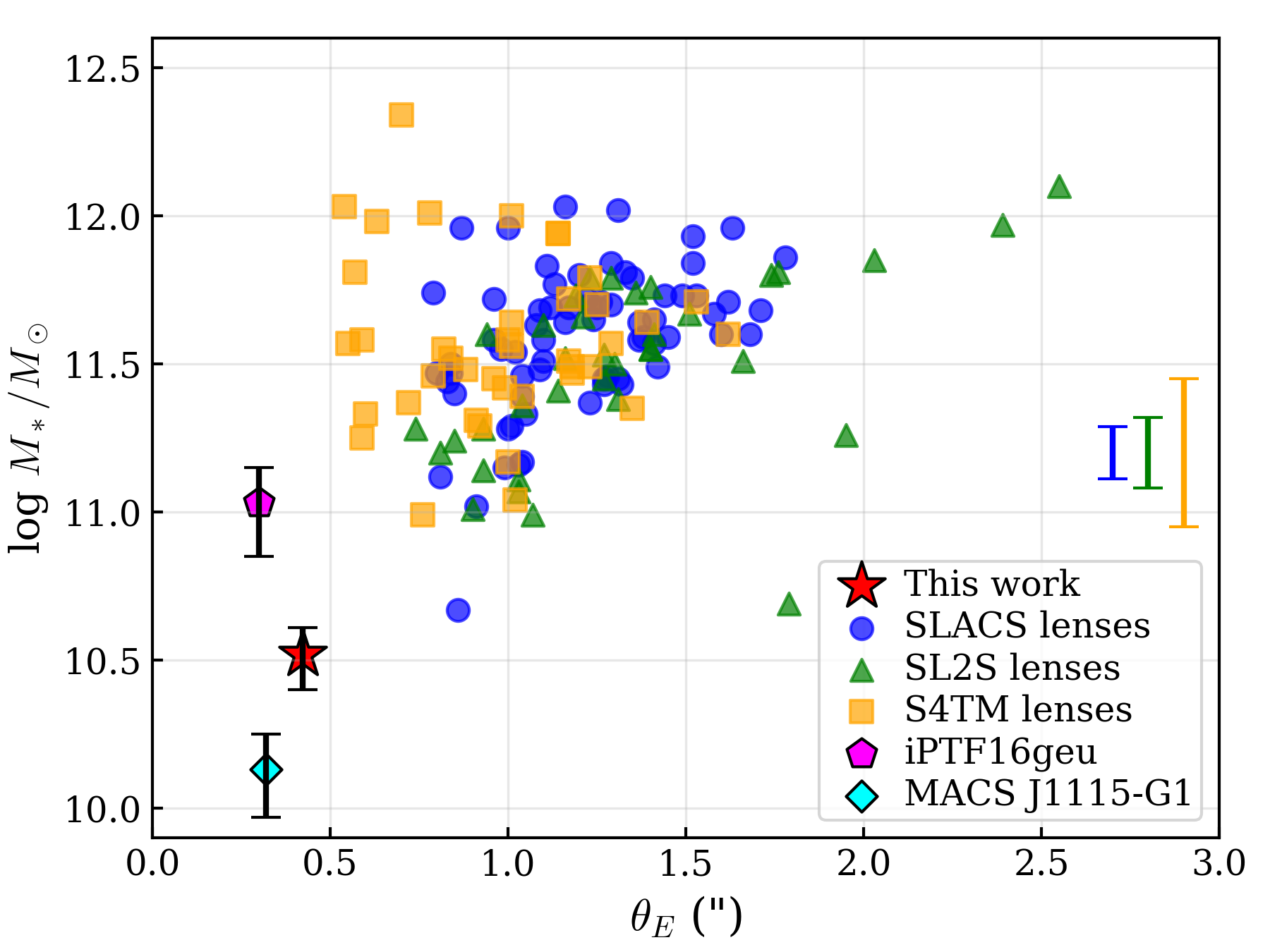}
    \caption{Stellar mass vs. Einstein radius for a representative subset of known strong lenses. Blue circles represent SLACS lenses, green triangles represent SL2S lenses, orange squares represent S4TM lenses, the magenta pentagon represents iPTF16geu, and the cyan diamond represents MACS\,J1115-G1. SMACS\,J0723-SL is indicated by the red star. The blue, green, and orange bars indicate the typical stellar mass uncertainties ($\pm 1\sigma$) for the SLACS, SL2S, and S4TM lenses.}
    \label{fig:distribution}
\end{figure}

The size, stellar mass, and stellar velocity dispersion measurements all suggest that the lens galaxy in SMACS\,J0723-SL is indeed a relatively low-mass early-type galaxy. To put into context, we constructed a compilation of previously known strong lenses from the literature. The compilation contains 148 lens galaxies at redshifts $\approx 0.05-0.9$ with published Einstein radius and stellar mass measurements, including 70 from the Sloan Lens ACS Survey (SLACS), 36 from the Strong Lenses in the Legacy Survey (SL2S), 40 from the SLACS for the Masses (S4TM) Survey, iPTF16geu, and another strong-lens galaxy discovered in the MACS\,J1115.9+0129 cluster (i.e., MACS\,J1115-G1). Although not complete, this compilation is representative of all currently known galaxy-scale strong lenses in terms of stellar mass and Einstein radius. For the SLACS lenses, Einstein radii and stellar masses were retrieved from \citet{2008ApJ...682..964B} and \citet{2009ApJ...705.1099A}. For the SL2S lenses, Einstein radii and stellar masses were retrieved from \citet{2013ApJ...777...97S}. For the S4TM lenses, Einstein radii and stellar masses were retrieved from \citet{2017ApJ...851...48S}. For iPTF16geu, we obtained its Einstein radius from \citet{2017Sci...356..291G} and its stellar mass from \citet{2025arXiv250101578A}. For MACS\,J1115-G1, Einstein radius and stellar mass were retrieved from \citet{2016MNRAS.458.1493P}. Figure~\ref{fig:distribution} shows the stellar mass-Einstein radius distributions for the known lens compilation\footnote{In the original papers, a Chabrier IMF \citep{2003PASP..115..763C} was assumed in the stellar mass estimations for the S4TM lenses, iPTF16geu, and MACS\,J1115-G1. To make a more fair comparison, in Figure~\ref{fig:distribution}, we added 0.25 dex to their reported stellar masses, which is a typical mass-to-light ratio offset between a Chabrier IMF and a Salpeter IMF for an old stellar population.} and SMACS\,J0723-SL. It becomes obvious that the lens galaxy in SMACS\,J0723-SL is substantially less massive than the majority of known lens galaxies. SMACS\,J0723-SL thus provides a rare opportunity to peek into a population of galaxies that is largely unexplored in previous strong-lensing studies.

It is worth making a brief comparison between SMACS\,J0723-SL and MACS\,J1115-G1, another low-mass strong-lens galaxy embedded in a galaxy cluster. According to \citet{2016MNRAS.458.1493P}, MACS\,J1115-G1 has a similar (lens) redshift of 0.353 and a similarly small Einstein radius of $0^{\prime \prime}.32 \pm 0^{\prime \prime}.04$. The stellar mass of MACS\,J1115-G1 is $(7.6 \pm 2.3) \times 10^{9} M_\odot$, assuming a Chabrier IMF (or $(1.4 \pm 0.4) \times 10^{10} M_\odot$ after converting to a Salpeter IMF), a factor of $\approx 2.4$ less massive than SMACS\,J0723-SL. On the other hand, the half-light radius of MACS\,J1115-G1 is 3.3 kpc, a factor of $\approx 3$ larger, suggesting that the stellar mass distribution of SMACS\,J0723-SL appears more compact. The projected dark-matter fraction within the half-light radius is close to 90\% for MACS\,J1115-G1, substantially higher than that of SMACS\,J0723-SL. More similar measurements will be extremely helpful in characterizing the internal mass structures of galaxies in this mass regime. 

SMACS\,J0723-SL is also a valuable discovery from the perspective of cluster lens modeling. Due to the intrinsic complexity, it is currently infeasible to model all cluster member galaxies in the same detail as is commonly done for galaxy-scale strong lenses. Instead, cluster member galaxies are usually assumed to follow the same mass distribution profile up to a normalization factor, which is determined through an empirical scaling relation \citep{1997MNRAS.287..833N, 2007ApJ...668..643L}. This conventional modeling approach may introduce significant systematics, especially as there is no consensus on the exact form of the mass profile or scaling relation \citep{2017MNRAS.472.3177M}. SMACS\,J0723-SL can provide an exclusive chance of validating the adopted mass profiles and scaling relations by comparing mass models from the conventional approach to those constructed with the strong-lensing features around SMACS\,J0723-SL taken into account (as achieved in this work). This type of test will provide important insights into dark-matter properties and cosmology that cluster lenses have been used to constrain. For example, \citet{Meneghetti2020} found that observed cluster member galaxies, based on the conventional modeling approach, have noticeably larger lensing cross sections than simulation predictions assuming cold dark matter. They suggested that this discrepancy could arise from the assumption about the nature of dark matter. It would be interesting to verify the estimation of lensing cross sections using systems like SMACS\,J0723-SL \citep[e.g.,][]{Granata2023}. 

Although the number of low-mass strong lenses ($M_* \sim 10^{10.5} M_\odot$) is very limited at the moment, ongoing and forthcoming facilities will soon increase the sample size by orders of magnitude. In particular, space missions such as Euclid and China Space Station Telescope (CSST) will be able to discover $10^5$ galaxy-scale strong lenses with thousands containing low-mass lens galaxies with Einstein radii as small as 0\farcs2 \citep[e.g.,][]{Collett2015, Cao2024}, both in clusters and in fields. With such a large sample, we expect to gain a far more comprehensive understanding of galaxy evolution, the nature of dark matter, and many other aspects.

\section{Conclusion} \label{sec:Conclusion}

In summary, we report the discovery of a rare, low-mass strong-lens system---SMACS\,J0723-SL in the \cluster cluster. MUSE spectroscopic data suggest that the lens and source galaxies are at redshifts of 0.3970 and 1.4792, respectively. By simultaneously modeling JWST NIRCam imaging data in six filters, i.e., F090W, F150W, F200W, F277W, F356W, and F444W, we find that the Einstein radius of the lens galaxy in SMACS\,J0723-SL is $0^{\prime \prime}.424 \pm 0^{\prime \prime}.012$, which corresponds to a total enclosed mass of $(3.7 \pm 0.2) \times 10^{10} M_\odot$. The half-light radius of the lens galaxy decreases from 0\farcs2604 (1.39 kpc) in the F090W to 0\farcs1935 (1.02 kpc) in F444W, and the central velocity dispersion is found to be $140 \pm 6$ km s$^{-1}$. The stellar mass of the lens galaxy is estimated to be $(3.3 \pm 0.8) \times 10^{10} M_\odot$ from fitting the photometry in six JWST filters. The dark-matter fraction within the half-light radius is $\approx (12 \pm 22) \%$. 

SMACS\,J0723-SL is one of the few known cases that contains a low-mass lens galaxy of $M_* \sim 10^{10.5} M_\odot$, which is about an order of magnitude lower than the average mass of all galaxy-scale strong lenses discovered to date. It hence presents a rare opportunity to investigate a population of galaxies that has remained largely unexplored in previous strong-lensing studies. In the near future, thousands of low-mass strong lenses with Einstein radii as small as 0\farcs2 will be discovered by missions such as Euclid and CSST, which will provide further constraints on galaxy evolution, dark matter, and cosmology.

\begin{acknowledgments}
We would like to thank the anonymous referee for helpful comments, which improved the presentation of this work. This work is supported by the National Key R\&D Program of China (grant No. 2023YFA1608100).
\end{acknowledgments}

%

\vspace{5mm}
\facilities{JWST, VLT:Yepun (MUSE)}


\software{lenstronomy \citep{2018PDU....22..189B}, pPXF \citep{2004PASP..116..138C, 2017MNRAS.466..798C, 2023MNRAS.526.3273C}, CIGALE \citep{2019A&A...622A.103B}
          }


\bibliography{reference}{}
\bibliographystyle{aasjournal}



\end{document}